# RATIONAL SECRET SHARING OVER AN ASYNCHRONOUS BROADCAST CHANNEL WITH INFORMATION THEORETIC SECURITY

William K. Moses Jr. and C. Pandu Rangan

Department of Computer Science and Engineering, Indian Institute of Technology
Madras, Chennai, India
`wkmjr3@gmail.com`
`prangan55@gmail.com`

## ABSTRACT

*We consider the problem of rational secret sharing introduced by Halpern and Teague [1], where the players involved in secret sharing play only if it is to their advantage. This can be characterized in the form of preferences. Players would prefer to get the secret than to not get it and secondly with lesser preference, they would like as few other players to get the secret as possible. Several positive results have already been published to efficiently solve the problem of rational secret sharing but only a handful of papers have touched upon the use of an asynchronous broadcast channel. [2] used cryptographic primitives, [3] used an interactive dealer, and [4] used an honest minority of players in order to handle an asynchronous broadcast channel.*

*In our paper, we propose an m-out-of-n rational secret sharing scheme which can function over an asynchronous broadcast channel without the use of cryptographic primitives and with a non-interactive dealer. This is possible because our scheme uses a small number, k+1, of honest players. The protocol is resilient to coalitions of size up to k and furthermore it is ε-resilient to coalitions of size up to and including m-1. The protocol will have a strict Nash equilibrium with probability $\Pr(\frac{k+1}{n})$ and an ε-Nash equilibrium with probability $\Pr(\frac{n-k-1}{n})$. Furthermore, our protocol is immune to backward induction.*

*Later on in the paper, we extend our results to include malicious players as well.*

*We also show that our protocol handles the possibility of a player deviating in order to force another player to get a wrong value in what we believe to be a more time efficient manner than was done in Asharov and Lindell [5].*

## 1. INTRODUCTION

### 1.1. Background

The classical problem of *m-out-of-n* secret sharing deals with distributing information about a secret to several players and then having them cooperate and work together in order to reconstruct that secret. More specifically, if *m* or more players come together, then no matter which players they are, they should be able to correctly reconstruct the secret. However, if less than *m* players come together, then they should not be able to reconstruct the secret. A solution to this problem came in the form of Shamir secret sharing [6]. Suppose we wanted it such that a minimum of at least *m* players must come together in order for a secret to be reconstructed. Then what we would do is construct a polynomial, $F(x)$, of degree *m-1* and make $F(0)$ the secret. We would then distribute a pair of values $(y, F(y))$ to each player where $y$ is a different value for each player. If at least $m$ players participate in the protocol, then by Lagrange's interpolation formula, they can reconstruct the secret.

The rational version of this classical problem alters the type of players involved in the game. In the classical version, players were either honest or malicious. However, in the rational version, players are rational, which means that they will only play the game if it is in their best interests to do so. In this scenario, we make a few assumptions about the players. First, that players prefer to learn the secret rather than not learn it. Secondly, that players prefer as few others as possible to learn the secret. With these two assumptions in mind, the problem ceases to be one of merely protecting honest players from the trickery of malicious players, and starts to be one of protecting every player from every other player and moreover, incentivizing every player to participate. With this new setting, we need a new solution concept in order to judge the effectiveness of protocols. This concept comes in the form of Nash equilibrium. A game is in Nash equilibrium if every player is playing her best response to every other player's best response.

In the rational setting, Shamir's scheme, which previously worked, proves unsuccessful as shown below. Suppose that $m^*$ players actually play an *m-out-of-n* Shamir secret sharing game.

**If $m^* < m$, then Nash equilibrium.** But the secret is not reconstructed.
**If $m^* = m$, then not a Nash equilibrium.** At this point, if a player decides not to send her share, then she ends up with everyone else's share and is able to reconstruct the secret while the others cannot.
**If $m^* > m$, then Nash equilibrium.** Even if a player does not send her share, the remaining players would. However, this is unstable because if enough people decide not to send their share, then this case degenerates to the previous case where $m^* = m$, which is not a Nash equilibrium.

Due to the unstable Nash equilibrium when $m^* > m$ and non-Nash equilibrium when $m^* = m$, it is ill advised to use Shamir's secret sharing as is in the rational setting. Several protocols have been devised to solve this problem [1,2,3,4,5,7,8,9,10,11,12,13], but only a handful have actually dealt with rational secret sharing using an asynchronous broadcast channel. These include Maleka et al.'s result [3] and Fuchsbauer et al.'s result [2], and Ong et al.'s result [4]. However, we contest that we can remove the interactive dealer [3] and cryptographic primitives [2] at the cost of assuming that a few of the participating players must be honest, but we believe that this is a reasonable assumption when the number is small, as in our scheme. This idea of assuming a small minority of honest players with a rational majority was first introduced by Ong et al. [4] and enabled them to obtain very strong results in rational secret sharing. They were able to also address the use of asynchronous broadcast channels, however their method differs from ours. Also, our protocol is able to handle coalitional deviations, which is one of the future directions of work mentioned in [4].

A rational secret sharing scheme consists of two algorithms, the dealer's algorithm and the player's algorithm. The dealer's algorithm is used to distribute shares to each of the players. In our scheme, the dealer's algorithm is only used once before the start of play. The player's algorithm is used by the players in order to interpret the information of their own shares as well as the information sent by other players and prescribes a set of actions to follow in order to progress the game. When we refer to the protocol in our paper, we are referring to the player's algorithm for the most part.

### 1.2. Our results

Our protocol is an *m-out-of-n* secret sharing scheme which utilizes an asynchronous broadcast channel, involves a non-interactive dealer, and does not use any cryptographic primitives.

Depending on the number of honest players *k+1* participating in the protocol, the protocol is resilient to coalitions of size up to *k* and furthermore it is ε-resilient to coalitions of size up to and including *m-1*. Choosing the right value of *k* really depends on how many honest players you believe to be active in the network and allows for a good tradeoff, more the number of players you believe play honestly, better the protection against coalitions.

In Asharov and Lindell's work [5], they talked about the concept of $U_i^f$-independence of any player *i* and it's impossibility in the case of synchronous and asynchronous networks. $U_i^f$ of a player refers to the utility gained by the player *i* when deviating in order to force another player to obtain a wrong value as the secret. $U_i^f$-dependence reflects how well a protocol deals with this utility of a player *i*. In order to handle $U_i^f$-dependence in non-simultaneous channels, they proposed a mechanism wherein they add a number of completely fake rounds to the protocol. In the case of 2-out-of-2 secret sharing using their mechanism, if the second player tries to deviate and achieve $U_i^f$ in a completely fake round, the first player will know and hence both players will achieve their respective $U_i^-$, i.e. the utility gained when a player does not get the secret. In this scenario, the second player stands to gain nothing from deviating in a completely fake round and the fact that she does not know which rounds are completely fake acts as a deterrent to her. The probability of deviating and fooling the other player becomes $\frac{E(r) - f}{E(r)}$, where *f* is the number of completely fake rounds and *E(r)* is the expected number of rounds (including the completely fake rounds). To achieve better protection, we need to increase the number of completely fake rounds in the protocol. This leads to a longer expected time. In our protocol, we can achieve the same sort of deterrent because despite a player's deviation, the honest players will always reveal the game to be real or fake. The only way that this will fail is if the deviating player manages to beat the authentication of the message. For information theoretic message authentication which uses $\log \frac{1}{\beta}$ bits, the probability of beating the authentication is β. By trading off between linear time and logarithmic number of bits, we are able to handle the $U_i^f$-dependence of any player *i* while reducing the expected running time of the protocol.

Furthermore, our protocol is immune to backward induction.

### 1.3. Organization

Section 2 discusses work related to this paper. Section 3 provides a background of the area with required definitions and assumptions made. Section 4 provides a detailed explanation and analysis of the proposed protocol. Section 5 extends our work to include malicious players. Section 6 wraps up the paper with some proposed lines of research to pursue.

## 2. RELATED WORK

Work has already been done in the area of asynchronous broadcast by [3], [2], and [4].

In [3], the authors suggested that by modifying their protocol and using repeated games with an interactive dealer, it was possible to have a working protocol for asynchronous broadcast. Our protocol makes use of repeated games, however it does not require an interactive dealer.

In [2], the authors proposed a working protocol for synchronous broadcast using cryptographic primitives and also extended their results to asynchronous broadcast as well by modifying their protocol. Our protocol does not require any cryptographic primitives but is instead information theoretically secure.

In [4], the authors proposed a working protocol for synchronous broadcast using the idea of an honest minority with rational majority and in the process obtained very good results both in terms of expected number of rounds taken to finish the protocol as well as in terms of the equilibrium used. They further extended this model to asynchronous broadcast, still maintaining good results. However, in their protocol, it is required that the subset of honest players is a random subset of $k = \omega(\log n)$, where *n*, the total number of players, is sufficiently large. Our protocol only places the restriction that $k < m < n$, where *k+1* players are honest in an *m-out-of-n* secret sharing scheme. The difference between the schemes is that in order for theirs to work properly, *n* must be sufficiently large, but ours will work for small groups of players as well as large groups. Furthermore, they gained an exact notion of equilibrium ($\varepsilon = 0$) at the cost of having an approximate notion of fairness, i.e. there is a negligible probability that the honest players may fail to compute the secret correctly if they follow the prescribed protocol. Our protocol however achieves an exact notion of fairness. Also, with regards to our equilibrium notion, we achieve ε-Nash equilibrium in most cases, but as the number of honest players involved in the reconstruction protocol increases, so does the probability of obtaining a strict Nash equilibrium. Finally, one of the future directions of work mentioned in [4] was to find a solution concept which would be resilient to coalitional deviations. An earlier version of their paper [14] showed coalition-proofness against stable coalitions in a model simpler than their fail-stop one. Our protocol is resilient to coalitions of size $< k$ and furthermore ε-resilient to coalitions of size up to and including *m-1*.

## 3. DEFINITIONS AND ASSUMPTIONS

In order to understand the work done in subsequent sections of this paper, we must first define a few important terms. Please note that definitions 3.1 to 3.7 are taken from Kol and Naor's paper [7] with slight modifications. Please note especially that the term game as used in their definitions has been changed to set of games in our definitions. This is done in order avoid confusion, due to the fact that we use repeated games in our paper.

### 3.1. Definitions

**Definition 3.1**: The utility function $u_i$ of a player is defined as a function which maps a player's actions to a payoff value in such a way that preferred actions result in higher payoff values.

**Definition 3.2**: We say that a player retrieves the designated value (the secret or *F(x)*) when outcome *o* is reached, if according to *o* the player quits and outputs the right value. Let *o* and *o'* be two possible outcomes of the set of games, and let *retrieve(o)* be the set of players retrieving the value when *o* is reached. If the following condition holds, then we say that the nature of the utility function is learning preferring: $u_i(o) > u_i(o')$ whenever $i \in retrieve(o)$ and $i \notin retrieve(o')$ (players prefer to learn).

**Definition 3.3**: Note that we call a vector of players' strategies a strategy profile, and use the following notations: $\alpha_{-i} = (\alpha_1, \ldots, \alpha_{i-1}, \alpha_{i+1}, \ldots, \alpha_n), (\alpha_{-i}, \alpha_i') = (\alpha_1, \ldots, \alpha_{i-1}, \alpha_i', \alpha_{i+1}, \ldots, \alpha_n)$ and $u_i(\sigma) = E_{o \sim O(\sigma)}[u_i(o)]$ where $O(\sigma)$ denotes the probability distribution over outcomes induced by the protocol σ. Now, a behavioural strategy profile σ for a protocol is said to be a Nash equilibrium if for every $i \in N$ and any behavioural strategy $\sigma_i'$, it holds that $u_i(\sigma_i, \sigma_{-i}) \geq u_i(\sigma_i', \sigma_{-i})$.

**Definition 3.4**: A behavioural strategy profile σ for a set of repeated games is said to be a strict Nash equilibrium if for every $i \in N$ and any behavioural strategy $\sigma_i'$, it holds that $u_i(\sigma_i, \sigma_{-i}) > u_i(\sigma_i', \sigma_{-i})$.

**Definition 3.5**: A behavioural strategy profile σ for a set of repeated games is said to be a ε-Nash equilibrium if for every $i \in N$ and any behavioural strategy $\sigma_i'$, it holds that $u_i(\sigma_i, \sigma_{-i}) + \varepsilon \geq u_i(\sigma_i', \sigma_{-i})$, where ε is a negligible value.

**Definition 3.6**: A coalition is a subset of the players who are active during the reconstruction phase.

**Definition 3.7**: A protocol is said to be resilient to coalitions of size *t* if even a coordinated deviation by a coalition of that size or less won't increase the utility of any of the players of the coalition. According to the prescribed strategy, each player of the coalition, $i \in C$, plays the strategy $\sigma_i$ and the remaining players play $\sigma_{-i}$. Let the player's deviating strategy be $\sigma_i^d$. Then the protocol is said to be resilient to coalitions of size *t* if $\forall i \in C, \forall |C| \leq t$, it holds that $u_i(\sigma_i, \sigma_{-i}) > u_i(\sigma_i^d, \sigma_{-i})$.

**Definition 3.8**: We say that a protocol is ε-resilient to a coalition of size *t* if no member of the coalition can gain more than ε in the process of a coordinated deviation by the coalition. According to the prescribed strategy, each player of the coalition, $i \in C$, plays the strategy $\sigma_i$ and the remaining players play $\sigma_{-i}$. Let the player's deviating strategy be $\sigma_i^d$. Then the protocol is said to be ε-resilient to coalitions of size *t* if $\forall i \in C, \forall |C| \leq t$, it holds that $u_i(\sigma_i, \sigma_{-i}) + \varepsilon \geq u_i(\sigma_i^d, \sigma_{-i})$, where ε is a negligible value.

**Definition 3.9**: We define the utility $U_i^f$ of a player *i* as the utility she gets if she is able to trick the other players into believing that a non-secret is the secret. $U_i^f$-dependence refers to how well the protocol deals with this utility for any player *i*.

**Definition 3.10**: Backward induction can be described as follows. If we know that the last round of a game is *r*, then we choose not to broadcast information in that round in order to gain in utility (if other players broadcast while we do not, then we gain the secret while others may not). Since each player thinks like this, common knowledge states that we all know that no one will broadcast in round *r*, and hence round *r-1* effectively becomes the last round. Similarly, everyone will not broadcast in *r-1* because it is now the last round and since that becomes common knowledge, round *r-2* becomes the last round. This type of thinking continues until at last we don't broadcast in any of the rounds. Similarly, if we are dealing with repeated games, as in this paper, if we know when the last game of the protocol will be played, then if we know when the last round (in our case, stage) of the last game is to be played, we can choose not to play that game and effectively the previous game becomes the last game.

**Definition 3.11**: In our protocol to reconstruct the secret, we use repeated games. In that context, we refer to the true game as the game, which upon completion, will reveal the actual secret.

### 3.2. Assumptions

As for the assumptions made, we follow several of those made in Fuchsbauer et al.'s paper [2] when it comes to an asynchronous broadcast channel.

1. We consider that an Asynchronous Broadcast Channel is present and connects all the *n* players such that a message sent from one player will be received by all the remaining players.
2. All players broadcast their values simultaneously.
3. Any message sent will eventually be received, even if it is at time ∞.
4. Rational and malicious players may schedule the message delivery. In other words, players who are not honest may schedule message delivery to benefit themselves.

## 4. OUR PROTOCOL

At the heart of our protocol is a 2-stage game which is repeated a number of times depending on a geometric distribution. First we describe the 2-stage game and then we describe how it is used.

The information shared in stage 1 and stage 2 of the game correspond to the 2 stages in [5]. In stage 1, we broadcast player *i*'s share of $R_j \otimes S_j$ where $R_j$ is a random number used in game *j* and $S_j$ is the possible secret used in game $j$. We also broadcast information theoretic authentication information about the stage 1 information as done in [7]. In stage 2, we broadcast player $i$'s share of $R_j$ as well as $i$'s share of a boolean indicator, which indicates whether the current game is the true game or not. Here as well, we broadcast information theoretic authentication information for stage 2. Note that we choose the value of $k$ according to the number of honest players "$k+1$" participating in the protocol.

The game is played as follows. Initially everyone broadcasts their share of $R_j \otimes S_j$ and it is reconstructed using *m-out-of-n* Shamir secret sharing. Once every player's share has been received and is verified, stage 2 commences. Now, every player broadcasts their share of $R_j$ and a boolean indicator. Both these values are reconstructed using *k-out-of-n* Shamir secret sharing. The value of $R_j \otimes S_j$ Xored with $R_j$ produces the possible secret and the reconstructed boolean indicator tells us whether this game was the true game or not. In stage 2, it is guaranteed that at least $k$ people broadcast (even though $k+1$ people are honest players, one of them may be the short player and this may be the true game, in which case $k$ other long players are required to reconstruct the secret). So, whether or not this game was the true game can be determined except in the very rare case that a player manages to break the information theoretic authentication check and pass off a forged stage 2 share. This would only happen with a negligible probability ε which can be lowered by increasing the bit size on the security checking "tag" and "hash" values.

This game is repeated $O(\frac{1}{\beta})$ times until the short player (the player who has less number of games to play) finishes playing all her games. At this point the short player should ideally broadcast ⊥, which would be a message understood by all parties to signify that the player has finished playing all her games. For our purposes, we may assume that ⊥ is represented by the value zero.

Table 1.  Dealer's share allotment protocol

---

**Dealer(y,$\beta$)**

Let $F = GF(p)$ where $p \geq |Y|$ prime, and associate each element of the secret set $Y$ with an element of $F$.

Denote by $G(\beta)$ the geometric distribution with parameter $\beta$.

- **Create the list of possible secrets and random numbers:**
    - Choose $l, d \sim G(\beta)$ such that $L = l + d - 1$ is the size of the full list of possible secrets.
    - Select a random ordering of the possible secrets such that the $l^{th}$ secret is the actual secret $y$.
    - Generate a list of size $L$ of random numbers.

- **Create shares:** Create *n* shares. One share will contain $l - 1$ cells and the remaining shares will contain $L$ cells. The player who gets the share with less number of cells is deemed the short player and the remaining players are deemed the long players. The values in each cell of the short player are the same as the values in the corresponding cells in the long players. The $l^{th}$ cell of the long players is considered the true game and the secret revealed after playing the game will be the real secret. Each cell corresponds to a 2-stage game and consists of the following information:

---

> *Stage 1 information:*
>
> - **Masked Secret:** An *m-out-of-n* Shamir share of $R_j \otimes S_j$, where $R_j$ is a randomly generated number for game $j$ and $S_j$, is a possible secret for game $j$.
> - **Authentication Information:** Information theoretic security checking as done in [7]. A "tag" to prove the authenticity of the masked secret and "hash values" to check the authenticity of other players' tags.
>
> *Stage 2 information:*
>
> - **Mask:** A *k-out-of-n* Shamir share of $R_j$.
> - **Indicator:** A *k-out-of-n* Shamir share of a boolean value indicating whether this game reveals the true secret or whether the game should be repeated.
> - **Authentication Information:** A "tag" to prove the authenticity of stage 2 information for this game and "hash values" to verify the authenticity of other players' tags.
>
> - **Add one more half cell (containing only stage 1 info) to each share.**
> - **Assign shares:** Randomly allot the shares to the players.

Table 2. Player *i*'s reconstruction protocol

> **Player$_i$(share)**
>
> Set $secret\_revealed \leftarrow FALSE$ and $cheater\_detected \leftarrow FALSE$.
>
> **Repeat the following until $secret\_revealed$ is $TRUE$ or $cheater\_detected$ is $TRUE$:**
>
> - **If the player's share ended (the player is at the final half cell containing only stage 1 info):**
>   - If this is stage 1 of the game:
>     - Broadcast the player's stage 1 tag and the player's share of $R_j \otimes S_j$.
>     - If anyone's message did not pass the authentication check, set $cheater\_detected \leftarrow TRUE$.
>   - If this is stage 2 of the game:
>     - Broadcast ⊥.
>     - If at least *k* people have broadcasted and their messages passed the authentication check, set $secret\_revealed \leftarrow TRUE$ and leave the game.
>     - If anyone's message did not pass the authentication check, set $cheater\_detected \leftarrow TRUE$.

- **If the player's share did not end:**
    - If this is stage 1 of the game:
        - Broadcast the player's stage 1 tag and the player's share of $R_j \otimes S_j$.
        - If anyone's message did not pass the authentication check, set $cheater\_detected \leftarrow TRUE$.
        - After the player has received all *n-1* of the other shares.
            - If they all passed the authentication check, proceed to stage 2.
            - Else leave the game.
    - If this is stage 2 of the game:
        - Broadcast the player's stage 2 tag and the player's share of the random number and indicator.
        - If anyone's message did not pass the authentication check, set $cheater\_detected \leftarrow TRUE$.
        - If at least *k-1* people have broadcasted and their messages passed the authentication check:
            - If the reconstructed indicator shows that this game revealed the true secret, then set $secret\_revealed \leftarrow TRUE$ and leave the game.
            - If the reconstructed indicator shows that this game did not reveal the true secret and someone broadcasted ⊥ in stage 1 or in stage 2, then set $cheater\_detected \leftarrow TRUE$ and leave the game.
        - After other messages have arrived, if all have passed authentication check, then proceed to next game in share. Else, leave the game.
- **Leave the game:** If $secret\_revealed$ is $TRUE$, then the secret can be reconstructed as follows. First reconstruct the masked secret using shares broadcasted in stage 1 of the game. Then construct the mask using shares broadcasted in stage 2 of the game. Xor the mask and the masked secret to get the secret. Quit and output the possible secret. If $secret\_revealed$ is $FALSE$, then the real secret was not obtained.

**Protocol Analysis**

We now analyze the protocol in some detail. First, we review the impact of using honest players. Then we move on to why we are able to obtain a strict Nash equilibrium and an ε-Nash equilibrium with given probabilities. Next we show why our protocol is immune to backward induction and finally we explain how the protocol handles the $U_i^f$-dependence of a player *i*.

The key to our protocol is the honest players. An honest player disregards her utilities and plays the game exactly as it is specified. Because of them, we can guarantee that some players will follow the protocol. Because of this assurance we have more control and can guarantee good results as was done in [4]. Also, if we look at previous results in the field of rational secret sharing, we come across [5], which basically said that if we make secret sharing a two stage process wherein we first share the Xor of a random number and the secret using *m-out-of-n* secret sharing and then follow that up by sharing the random number using *l-out-of-n* secret sharing, where *l* is any number less than *m*, then we could provide an incentive to players to not act maliciously, because even if they did, they would never benefit from it. This was analyzed for simultaneous secret sharing and proven in the paper. However, in the cases of synchronous broadcast and asynchronous broadcast, the incentive falls through and the idea cannot be used as is. [4] used another approach coupled with a small minority of honest players in order to obtain good results for synchronous broadcast. By using the idea of [5] coupled with honest players, we are able to obtain results for asynchronous broadcast.

As to our protocol having a strict Nash equilibrium with probability $\Pr(\frac{k+1}{n})$ and an ε-Nash equilibrium with probability $\Pr(\frac{n-k-1}{n})$, the reason for this rests solely with who the short player is. When the short player is honest, we can guarantee that she will not join a coalition or try to forge a fake secret during the true game. In that case, due to the way in which we choose $\beta$, it is strictly better for all the players involved to follow the protocol than to deviate. However, if the short player is not an honest player, then she may try to forge a secret and may succeed with a negligible probability of ε. Hence, when the short player is honest, we have a strict Nash equilibrium, else we have an ε-Nash equilibrium. The short player will be an honest player with a probability of $\Pr(\frac{k+1}{n})$.

Our protocol is immune to backward induction because it maintains the property that after any history that can be reached by the protocol, the protocol is still a strict Nash equilibrium with probability $\Pr(\frac{k+1}{n})$ and an ε-Nash equilibrium with probability $\Pr(\frac{n-k-1}{n})$. This is because one player is short while the others are long. No one knows if they are the short player or the long player and hence they are forced to keep playing. Furthermore, the honest players will always play irregardless of utility.

An idea also discussed in [5] is that of the $U_i^f$-dependence of a player *i*. It basically deals with the fact that a player may have something to gain by forcing other players to think they have the correct secret when in fact they don't, essentially making those players obtain a *fake secret*, and asks if it is possible to create a protocol where the desire to force others to obtain such a fake secret may be counteracted. Our protocol deals with this by intrinsically assuming that a possible secret is false until proven true. That is to say that even if a party *i* aborts early in order to achieve a gain of $U_i^f$, we don't assume the secret has been gained until and unless the secret is proven to be the true secret by the honest players, using a boolean indicator. By the same token, one might consider the possibility of a player, who somehow manages to discover which game reveals the true secret, trying to make it appear to be a fake secret by sending an incorrect message. Our protocol is designed to detect this trickery using information theoretic authentication and will correctly determine whether the current possible secret is indeed the true secret with a negligible error probability of ε, where the information theoretic authentication of the messages uses $\log \frac{1}{\beta}$ bits and $\beta$ is the parameter used for the geometric distribution in the dealer's protocol.

**The values of $\beta_0$ and $c_0$:**

We now need to calculate two values before we proceed to formulate our theorem. This analysis is present in Kol and Naor's paper [7] and will be repeated below (with slight tweaks) for the sake of understanding. We first define the utility values for a player $i$ as follows:

- $U_i$ is the utility of a player $i$ when she obtains the secret along with at least one player not belonging to any coalition of which she is a part.

- $U_i^+$ is the utility of a player $i$ when she obtains the secret and no player, other than those belonging to her coalition, obtains the secret.

- $U_i^-$ is the utility of a player $i$ when she does not obtain the secret.

Now, if the set of secrets $Y$ follows a distribution $D$, then let us assume that $b \in Y$ is the secret with the highest probability of being the actual secret according to $D$. In other words, $\forall x \in Y, D(b) \geq D(x)$. Let the probability that a player $i$ can guess the secret given the distribution $D$ and her share be $g$.

If the player follows the protocol, then she will get $U_i$. If she deviates and gets the secret, then she will get a utility of $U_i^+$. If she deviates but does not get the secret, with a probability of $1 - g$, then she stands to get a utility of $U_i^-$. Now, it matters to us to ensure that the expected utility from following the protocol is more than the expected utility from deviating from the protocol. So, the following must hold true:

*Expected utility (deviating) < Expected utility (following protocol)*

$$g * U_i^+ + (1 - g) * U_i^- < U_i$$

$$g * (U_i^+ - U_i^-) < U_i - U_i^-$$

$$g < \frac{U_i - U_i^-}{(U_i^+ - U_i^-)}$$

Let us call the ratio $\frac{U_i - U_i^-}{(U_i^+ - U_i^-)}$ for a player $i$, $c_i$. Since every player has learning preferring utilities, it follows that $U_i^- < U_i \leq U_i^+$ and hence $c_i > 0$ for every player $i$. So, for a given player $i$ to follow the protocol, we must ensure that the chance of her deviating and getting the secret is less than $c_i$. In order for the protocol to work for every player, we must ensure that the probability of deviating and getting the correct secret is less than all players' $c_i$'s. For a set of players $N$, we require

$$g < min_{i \in N} c_i$$

We call the value $min_{i \in N} c_i$ as $c_0$. Since every player's guess is at least as good as $D(b)$ we should require $D(b) < c_0$.

In our theorem we use $\beta_0$ to cap the value of β, which is the parameter of the geometric distribution used in the dealer's algorithm. For now, we state that the value of $\beta_0$ is $min_{i \in N} \left\{ \frac{c_i - D(b)}{c_i - D(b) + 2*z*n + 1} \right\}$ where $z = |Y|$. The reason for this value lies in the proof of Theorem 1. However, do note that since we require $c_0 > D(b)$, it follows that $c_i > D(b)$ which implies that $\beta_0 > 0$ and hence this is a valid cap of $\beta$.

**Theorem 1.** *Let $Y$ be a finite set of secrets with distribution $D$, and let each rational player have learning preferring utilities. If $D(b) < c_0$, then for $\beta < \beta_0$ and for all $2 \leq m \leq n$, the scheme described above is, for Y,*

- *an asynchronous strict rational m-out-of-n secret sharing scheme with probability $Pr(\frac{k+1}{n})$,*

- *an asynchronous ε-rational m-out-of-n secret sharing scheme with probability $Pr(\frac{n-k-1}{n})$,*
- *immune to backward induction,*
- *and handles $U_i^f$-dependence of any player i in a time efficient manner.*

*Depending upon the number of honest players k+1, where $2 \leq k \leq m-1$, participating in the k-out-of-n secret sharing stage of each game, the protocol will be resilient to coalitions of size $< k$ and furthermore ε-resilient to coalitions of size up to and including m-1. The scheme has expected round complexity of $O\left(\frac{1}{\beta}\right)$ and expected share size of $O\left(\frac{1}{\beta}\log\frac{1}{\beta}\right)$.*

**Proof** In order to prove this theorem, we need to prove the following things.

1. The expected round complexity of the reconstruction protocol and the expected share sizes are as claimed.

2. Any group of *m-1* players or less will not be able to learn anything about the secret before the game starts.

3. Any group of *m* players or more, where *k+1* of them are honest, correctly following the protocol will be able to reconstruct the secret.

4. The strategies prescribed to every subset of *m* or more players, where *k+1* of them are honest, will be strict best responses with a probability of $Pr(\frac{k+1}{n})$ and will be ε-best responses with a probability of $Pr(\frac{n-k-1}{n})$.

5. The protocol is resilient to coalitions of size $< k$ and ε-resilient to coalitions of size up to and including *m-1*.

6. The protocol is immune to backward induction.

7. The proposed scheme handles $U_i^f$-dependence of any player *i* in a time efficient manner.

To show (1), we can see that the total number of games is chosen according to the geometric distribution $G(\beta)$ and hence the expected round complexity is $O\left(\frac{1}{\beta}\right)$. Also, the size of each cell of a share pertaining to a game will be $O\left(\log\frac{1}{\beta}\right)$. This is due to the information theoretic authentication, which must be able to prevent players from simply guessing the "tag" values with a probability greater than $\beta$ and as such trying to forge their own "tag" values for fake information. Therefore the expected size of each share will be $O\left(\frac{1}{\beta}\log\frac{1}{\beta}\right)$.

To show (2), we note that the first stage of each game played requires *m* players to come together in order to reconstruct $R_j \otimes S_j$, hence any group of less than *m* players will not be able to learn anything about the secret prior to the start of the game.

For (3), we claim that if *m* or more players correctly follow our protocol, then they will be able to reconstruct the secret after the true game.

To show (4), we need to show that as long as no deviation was encountered, every player that does not know the secret is strictly better off with a probability of $Pr(\frac{k+1}{n})$ and ε-better off with a probability of $Pr(\frac{n-k-1}{n})$ to follow the protocol. This is quite similar to what was proven in Kol and Naor's paper [7] for Theorem 5.2. However, in our case, we are using an asynchronous channel instead of the simultaneous channel that they used. As such, our proof also focuses on showing how our protocol tackles the asynchrony present in the system.

We will first show that every player that does not know the secret is strictly better off with a probability of $Pr\left(\frac{k+1}{n}\right)$ and ε-better off with a probability of $Pr\left(\frac{n-k-1}{n}\right)$ to follow the protocol. Following this, we show how our protocol tackles the issue of asynchrony.

Let us denote the current game by $t$ and let $s_i$ be the size of the share assigned to player $i$. Let $b$ be the value with the highest probability of being the secret according to the probability distribution and let $z = |Y|$. There are 2 cases to be considered, which are as follows.

**Case 1: Player $i$'s share has ended ($t = s_i + 1$)**

In this case, player $i$ knows that she is the short player. One of two things will be true:

(i) Player $i$ is an honest player. This is possible with a probability $Pr\left(\frac{k+1}{n}\right)$.

If this is the case, then she will not deviate from the protocol and the protocol will be a strict rational protocol.

(ii) Player $i$ is a rational player. This is possible with probability $Pr\left(\frac{n-k-1}{n}\right)$.

In this case, she will try to deviate and achieve a utility of $U^+$. This is only possible if she is able to trick others into believing that this game is not the true game, by modifying the indicator and bypassing the authentication. She may do so with a probability of $\beta$. Let the probability of getting the secret by deviating be $\alpha$. It follows that

$$\alpha' = \beta$$

It is enough to require that

$$\alpha' < c_i$$
$$\beta < c_i$$
$$\beta < \beta_0$$

**Case 2: Player $i$'s share has not ended ($1 \leq t \leq s_i$)**

The part of the proof for this case is the same as given in Kol and Naor's paper [7]. A copy of the proof presented for this case for Theorem 5.2 in Kol and Naor's paper [7] is given in the appendix (with slight tweaks). At the end of it, we see that even in this case it is enough to require $\beta < \beta_0$ in order to deter players from deviating. Hence in both cases, the players will follow the protocol so long as $\beta < \beta_0$ given that $D(b) < c_0$.

Furthermore, our protocol also effectively handles asynchronous broadcast, and this may be shown as follows. In the case of [7], the proof as it was presented was sufficient for a simultaneous broadcast channel. In the case of asynchronous broadcast, several new problems arise which must be handled. The problems are:

1. Ordering the messages of a players' shares so that other players knows which messages to combine together. (Figuring out which of the several games' data of another player's share has been received. And then making sure that all that all the data being used to reconstruct the secret belongs to the same game.)

2. Synchronizing the players to prevent any player from gaining an unfair advantage. (Suppose one player chooses not to send messages. How do we distinguish this case from that of a player's message taking infinitely long to reach.)

The problems are addressed in our protocol as follows:

1. The problem of ordering of messages was taken care of thanks to the requirements for players to move to the different stages of a game. For a player to move from stage 1 of a game to stage 2 of the same game, she must first wait for the messages from **all** the

    other players. Furthermore, all the messages must pass the information theoretic authentication check. Only then can she move on to stage 2. Since the crucial information required to reconstruct the secret is transmitted in the second stage of the game, it follows that players will broadcast the correct required information in stage 1. Now, if the game currently being played is the true game, then it suffices for the honest players to play as they should, even if everyone else keeps silent. And since any message broadcasted will definitely reach (we had assumed this earlier just as Fuchsbauer et al.'s [2] assumed it in their paper), we are guaranteed that the protocol succeeds in having everyone learn the secret. However, should the current game not be the true game, then the incentive for all the players to transmit their information is that if they do not, then the honest players will not proceed to the next game and then there is no guarantee that the players will eventually discover the secret. A player may try to guess which game is the correct game and then deviate in the final step, but deviation is not desired since we have proved that by capping $\beta$ with $\beta_0$, player's are suitably deterred from deviating from the given protocol.

2. Synchronization of players is implemented by forcing players to play games one by one, and furthermore, each game stage by stage. There is a clear end to each game and to each stage and this provides us with a means of synchronization. Furthermore, players have no reason to deviate by not sending messages. This is taken care of because $\beta < \beta_0$.

To show (5), it is clear that as long as $k+1$ honest players play the game, $R_j$ will always be shared with the other players and hence the secret will be reconstructed. Furthermore, if the number of players in the coalition exceeds $k-1$ but is less than $m$, then the coalition can only succeed if one of its members successfully cheats the authentication and makes others believe that the true game is not the true game. This is only possible with a negligible probability of ε.

For (6), consider the following. The long players do not know which game is the final game until after it has occurred. As such, they won't know when to deviate until it is too late. Only the short player has a notion of which game is the last game, but that only occurs after her share has ended. Hence, the protocol is immune to backward induction.

In order to prove (7), we note that a player $i$ can only achieve a utility of $U_i^f$ if she is able to trick other players into obtaining an incorrect secret. The flip side is that if a player deviates, but the remaining players are not tricked, then the player does not obtain a utility of $U_i^f$. In our protocol, even if a player deviates (by say broadcasting wrong values in stage 1 or stage 2 of a game), unless she is able to pass the information theoretic authentication check, which is possible with only a negligible probability of $\beta$. Furthermore, for the same reason (authentication check) she cannot force others to obtain a wrong value for the secret by faking the true game, since this is possible with only a negligible probability of $\beta$ and also because the honest players will always act properly and by following the protocol, will alert others to any deception via the indicator. Hence the issue of $U_i^f$-dependence is addressed via the strength of the information theoretic checking. By increasing the strength of the checking (i.e. increasing the number of bits required for the checking and thereby increasing the cell size), we can greatly reduce the $U_i^f$-dependence to the point where only if $U_i^f \gg U_i$, will the player attempt deviating in this manner. This can be done without having to increase the number of games played, compared to the solution used in [5].

□

## 5. ADDITION OF MALICIOUS PLAYERS

A malicious player is someone who disregards her utility values and whose only goal is to disrupt the game. They can do this by causing the game to stop early, misleading others to believe that they have the right secret when they don't, or causing some players to get the secret and others not to. A step in the direction of dealing with malicious players was made by Lysyanskaya and Triandopoulos in [8], where they dealt with a situation where both rational and malicious players were involved in a game. They concluded that it was possible to play such a game but they were unable to prevent early stoppage. Our protocol is also able to function properly even with malicious players, but it cannot prevent early stoppage. It can, however, prevent the other two problems mentioned above. However, **Theorem 1.** needs to be modified in order to incorporate this new element.

**Theorem 2.** *Let Y be a finite set of secrets with distribution D, and let each rational player have learning preferring utilities. If $D(b) < c_0$, then for $\beta < \beta_0$ and for all $2 \leq m \leq n$, the scheme described above is, for Y,*

- *an asynchronous strict rational m-out-of-n secret sharing scheme with probability $Pr\left(\frac{k+1}{n}\right)$,*
- *an asynchronous ε-rational m-out-of-n secret sharing scheme with probability $Pr\left(\frac{n-k-1}{n}\right)$,*
- *immune to backward induction,*
- *and handles $U_i^f$ -dependence of any player i in a time efficient manner.*

*Let t players be the number of malicious players actively involved in the protocol. Depending upon the number of honest players k+1, where $2 \leq k \leq m-1$, participating in the k-out-of-n secret sharing stage of each game, the protocol will be resilient to coalitions of size $< k - t$ and ε-resilient to coalitions of size up to and including m - 1 - t. The scheme has expected round complexity of $O\left(\frac{1}{\beta}\right)$ and expected share size of $O\left(\frac{1}{\beta} \log \frac{1}{\beta}\right)$.*

This change in the theorem occurs because one way the malicious players can disrupt the game is by sending their shares to one player, through a side channel, but not to others. In order to avoid this, we must reduce the size of coalitions so that even if all malicious players send their shares to members of a coalition, that coalition will still not be able to reconstruct the secret on their own.

The proof for this theorem follows in the same vein as that of Theorem 1.

## 6. CONCLUSION AND FUTURE WORK

In this paper, we have proposed an *m-out-of-n* RSS protocol for the case of asynchronous broadcast, which makes use of a small number of honest players in order to achieve information theoretic security and protect against coalitions to a given extent. Further directions of work include:

- Probing the case of asynchronous point to point communication through the lens of information theoretic security.
- Further improving the coalition resilience for the asynchronous broadcast scenario.
- Further improving the communication complexity for the asynchronous broadcast scenario.

## A. PROOF OF CASE 2 (PLAYER $i$'S SHARE HAS NOT ENDED)

The following is the proof as to why any player $i$ is better off following the prescribed strategy when her share has not ended, i.e. $1 \leq t \leq s_i$, as long as $\beta < \beta_0$. This proof is taken from Kol and Naor's paper [7] as the proof for case 2 under the 4[th] point for the proof of Theorem 5.2. The proof has been tweaked to suit the interests of the protocol in the current paper.

We denote by *true_game_prob* the probability that the current game is the true game, given the player's share and transcript of moves thus far. The idea now is to prove that *true_game_prob* is a very small value. Intuitively, if a player is near the end of her share, then she believes she has the short share, and if the end of her share is still far far away, then she assumes that she is the long player and thus any future game could be the true game.

**Claim 1.** $true\_game\_prob \leq \frac{z*n*\beta}{1-\beta}$

**Proof of Claim 1.** The only parameters viewed by player $i$ that are relevant when determining whether the current game is the true game are: the number of the current game, $t$, player $i$'s share size, $s_i$, and the current unmasked possible secret $y_t$ (player $i$ learned this secret after the $t-1^{th}$ game. The other values viewed by player $i$ are independent of the secret and its revelation time. Assume that $s_i = k$ $(k > t), y_t = a$, and that player $i_0$ is the one with the short share.

$$true\_game\_prob = \Pr(l = t | l \geq t \wedge y_t = a \wedge s_i = k)$$

$$= \frac{\Pr(l = t \wedge y_t = a \wedge s_i = k)}{\Pr(l \geq t \wedge y_t = a \wedge s_i = k)}$$

$$= \frac{\Pr(i \neq i_0 \wedge l = t \wedge y_t = a \wedge s_i = k)}{\Pr(i = i_0 \wedge l \geq t \wedge y_t = a \wedge s_i = k) + \Pr(i \neq i_0 \wedge l \geq t \wedge y_t = a \wedge s_i = k)}$$

We now calculate the individual values of the numerator and denominator of the last fraction. Recall that if $t$ is the true game ($t = l$) then $y_t = y$, where $y$ is the real secret, otherwise $y_t = r_t$ for a randomly chosen $r_t \in Y$.

The term in the numerator:

$$\Pr(i \neq i_0 \wedge l = t \wedge y_t = a \wedge s_i = k)$$

$$= \Pr(i \neq i_0) * \Pr(l = t) * \Pr(y = a) * \Pr(d = k - t + 1)$$

$$= \frac{n-1}{n} * \beta * (1-\beta)^{t-1} * D(a) * \beta * (1-\beta)^{k-t}$$

$$= \frac{n-1}{n} * D(a) * \beta^2 * (1-\beta)^{k-1}$$

The first term in the denominator:

$$\Pr(i = i_0 \wedge l \geq t \wedge y_t = a \wedge s_i = k)$$

$$= \Pr(i = i_0) * \Pr(r_t = a) * \Pr(l = k + 1)$$

$$= \frac{1}{n} * \frac{1}{z} * \beta * (1-\beta)^k$$

The second term in the denominator:

$$\Pr(i \neq i_0 \wedge l \geq t \wedge y_t = a \wedge s_i = k)$$

$$= \Pr(i \neq i_0) * \Pr(r_t = a) * \sum_{j=t+1}^{k} \Pr(l = j) * \Pr(d = k - j + 1)$$

($t$ is not the true game) $+$

$$\Pr(i \neq i_0) * \Pr(y = a) * \Pr(l = t) * \Pr(d = k - t + 1)$$

($t$ is the true game)

$$= \frac{n-1}{n} * \frac{1}{z} * \sum_{j=t+1}^{k} \beta * (1-\beta)^{j-1} * \beta * (1-\beta)^{k-j} +$$

$$\frac{n-1}{n} * D(a) * \beta * (1-\beta)^{t-1} * \beta * (1-\beta)^{k-t}$$

$$= \frac{n-1}{n} * \beta^2 * (1-\beta)^{k-1} * \left(\frac{1}{z} * (k-t) + D(a)\right)$$

Therefore,

$$true\_game\_prob$$

$$= \frac{\frac{n-1}{n} * D(a) * \beta^2 * (1-\beta)^{k-1}}{\frac{1}{n} * \frac{1}{z} * \beta * (1-\beta)^k + \frac{n-1}{n} * \beta^2 * (1-\beta)^{k-1} * (\frac{1}{z} * (k-t) + D(a))}$$

$$= \frac{(n-1) * D(a) * \beta}{\frac{1}{z} * (1-\beta) + (n-1) * \beta * (\frac{1}{z} * (k-t) + D(a))}$$

$$\leq \frac{z * n * \beta}{1 - \beta}$$

♦

After completing some number of games in her share, the distribution over the secrets for player $i$ may have changed. Let us call this new distribution $D'$. $D'$ differs from $D$ because when the current unmasked secret is $y_t$, the probability of $y_t$ being the real secret increases. Corresponding to this new distribution, let the element with the highest chance of being the secret be $b' \in Y$. As we can infer, the higher the $D'(b')$, the better chance player $i$ has of guessing the secret correctly.

**Claim 2.** $D'(b') \leq true\_game\_prob + D(b)$

**Proof of Claim 2.**

We see that $D'(b')$ will have its highest value when the current unmasked secret is the value which had the highest chance of being the real secret according to the initial distribution, i.e. $y_t = b$. When this is the case, automatically $b = b'$. Therefore:

$$D'(b') \leq \Pr(y = b | l \geq t \wedge y_t = b \wedge s_i = k)$$

$$= \Pr(y = b \wedge l = t | l \geq t \wedge y_t = b \wedge s_i = k) +$$

$$\Pr(y = b \wedge l \neq t | l \geq t \wedge y_t = b \wedge s_i = k)$$

$$\leq true\_game\_prob + D(b)$$

♦

The final part of the proof comes from the fact that if a player $i$ deviates in the current game, then only if one of 3 conditions occurs, can she get the secret. These 3 conditions are:

- The current game is the true game. This occurs with probability $true\_game\_prob$.

- In case this is not the true game, then player $i$ was not caught in the act of cheating. This is true only if she beats the information theoretic authentication check and this occurs with a probability of $\beta$.

- Finally, in case player $i$ was caught, then she was able to guess the correct secret. The player's best chance of guessing the secret is $D'(b')$.

Once again, we require that the probability of deviating but still getting the secret of a player $i$ be capped by $c_i$. Hence:

$$true\_game\_prob + \beta + D'(b') < c_i$$

$$2 * true\_game\_prob + \beta + D(b) < c_i$$

$$\frac{2 * z * n * \beta}{1 - \beta} + \beta < c_i - D(b)$$

$$2 * z * n * \beta + (1 - \beta) * \beta < (1 - \beta) * (c_i - D(b))$$

$$\beta * (2 * z * n + 1 + c_i - D(b)) < c_i - D(b)$$

$$\beta < \frac{c_i - D(b)}{c_i - D(b) + 2 * z * n + 1}$$

$$\beta < \beta_0$$

∎

**Authors**

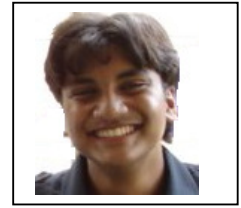

**William K. Moses, Jr.**

He is currently pursuing a Masters degree in the Dept. of Computer Science and Engineering of IIT Madras. His current research focuses on Rational Cryptography.

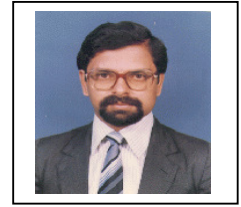

**C. Pandu Rangan**

He is currently a Professor in the Dept. of Computer Science and Engineering of IIT Madras. His research focuses on the design of pragmatic algorithms. His research interests include 1) Restricting the problem domain 2) Approximate algorithm design 3) Randomized algorithms 4) Parallel and VLSI algorithms and 5) Cryptography Applications.